\documentclass[graybox]{svmult}

\usepackage{amssymb}
\usepackage{mathptmx}       
\usepackage{helvet}         
\usepackage{courier}        
\usepackage{type1cm}        
\usepackage{makeidx}         
\usepackage{graphicx}        
\usepackage{multicol}        
\usepackage[bottom]{footmisc}
\usepackage{latexsym, amsfonts, mathcomp}
\makeindex             

\newcommand{\sun}{$_{\odot}$}

\newcommand       \apj          {ApJ}
\newcommand       \apjl         {ApJL}
\newcommand       \aap          {A\&A}
\newcommand       \nat          {Nature}
\newcommand       \mnras        {MNRAS}

\newcommand       \aj      {AJ}

\newcommand       \pasj   {PASJ}
\newcommand      \apjs {ApJ Supplements}

\newcommand      \bain {Bulletin Astronomical Institutes of the
  Netherlands}
\newcommand    \pasa {PASA - Publications of the Astronomical Society of Australia}

\begin{document}

\title*{X-Ray Binaries}
\author{J. Casares \and P.G. Jonker \and G. Israelian }
\institute{ J. Casares \at
  Instituto de Astrof\'\i{}sica de Canarias,  E--38205 La Laguna, S/C de Tenerife, Spain.\\
  Departamento de Astrof\'\i{}sica, Universidad de La Laguna, E--38206
  La Laguna, S/C de Tenerife, Spain. \\
 Department of Physics, Astrophysics,
   University of Oxford, Keble Road, Oxford OX1 3RH, UK  \email{jorge.casares@iac.es}
\and P.G. Jonker \at SRON, Netherlands Institute for Space Research,
  Sorbonnelaan 2,
  3584~CA, Utrecht, The Netherlands.\\
  Department of Astrophysics/IMAPP, Radboud University Nijmegen,
  P.O.~Box 9010, 6500 GL, Nijmegen, The Netherlands.\\
  \email{P.Jonker@sron.nl} 
\and G. Israelian \at
  Instituto de Astrof\'\i{}sica de Canarias,  E--38205 La Laguna, S/C de Tenerife, Spain.\\
  Departamento de Astrof\'\i{}sica, Universidad de La Laguna, E--38206
  La Laguna, S/C de Tenerife, Spain.  \email{gil@iac.es} }
\maketitle

\abstract{ This chapter discusses the implications of X-ray binaries
  on our knowledge of Type Ibc and Type II supernovae.  X-ray binaries
  contain accreting neutron stars and stellar--mass black holes which
  are the end points of massive star evolution. Studying these
  remnants thus provides clues to understanding the evolutionary
  processes that lead to their formation. We focus here on the
  distributions of dynamical masses, space velocities and chemical
  anomalies of their companion stars. These three observational
  features provide unique information on the physics of core collapse
  and supernovae explosions within interacting binary systems.  There
  is suggestive evidence for a gap between $\approx$2-5 M\sun~
 in the observed mass distribution.  This might be related to the physics
  of the supernova explosions although selections effects and possible
  systematics may be important.  The difference between neutron star mass
  measurements in low-mass X-ray binaries (LMXBs) and pulsar 
  masses in high-mass X-ray binaries (HMXBs) reflect their
  different accretion histories, with the latter presenting values
  close to birth masses. On the other hand, black holes in LMXBs
  appear to be limited to $\lesssim$12 M\sun~because of strong
  mass-loss during the wind Wolf-Rayet phase.  Detailed studies of a
  limited sample of black-hole X-ray binaries suggest that the more
  massive black holes have a lower space velocity, which could be 
  explained if they formed through direct collapse. 
 Conversely, the formation of low-mass black holes through a supernova
  explosion implies that large escape velocities are possible through
  ensuing natal and/or Blaauw kicks. Finally, chemical abundance studies of
  the companion stars in seven X-ray binaries indicate they are
  metal-rich (all except GRO J1655-40) and possess large peculiar
  abundances of $\alpha$-elements. Comparison with supernova models
  is, however, not straightforward given current uncertainties in
  model parameters such as mixing.
  \\
}

\section{Introduction}
\label{sec:1}
\begin{figure}[h] \hbox{\hspace{+2.0cm}
\vspace{0.1cm}
\includegraphics[angle=0,width=7cm]{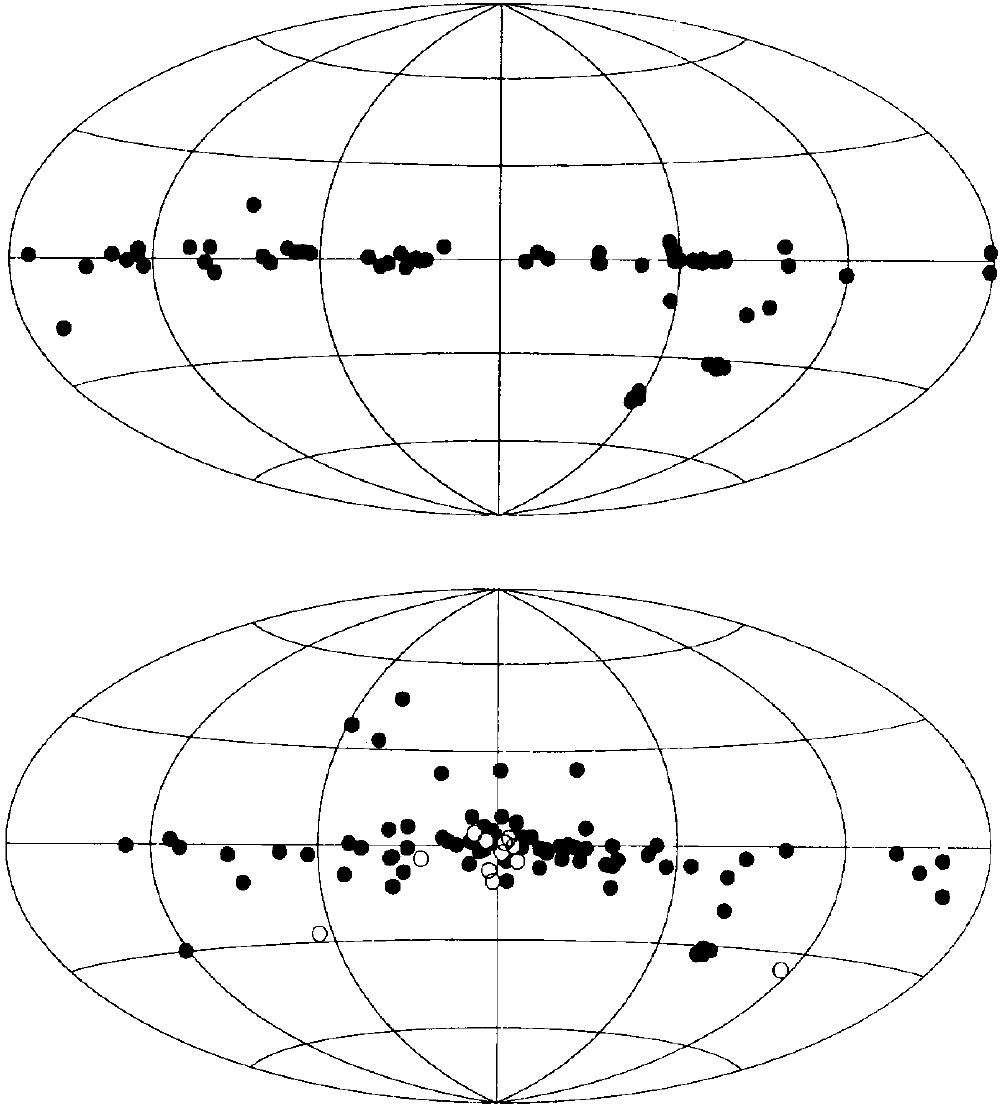}
\hspace{1.0cm}} \caption{
Galactic distribution of HMXBs (top) and LMXBs (bottom). Open circles indicate LMXBs in Globular Clusters. 
From \cite{vanparadijs98}.}
\label{fig:1}       
\end{figure}

X-ray binaries contain compact stellar remnants
accreting from "normal" companion stars. Therefore, they provide ideal
opportunities for probing the core-collapse of
massive stars in a binary environment and are thus able to constrain
the physics of Type Ibc and Type II
supernovae. These compact remnants are revealed by
persistent/transient X-ray activity which is triggered by mass
accretion. Observationally, they come in three
flavors -- pulsars, neutron stars
and black holes -- that are paired with companion
(donor) stars of a wide range of masses.  Historically, X-ray binaries
have been classified according to the donor mass as either 
Low Mass X-ray Binaries (LMXBs) or 
High Mass X-ray Binaries (HMXBs). The former are
fueled by accretion discs supplied by a $\lesssim$1 M\sun~Roche-lobe
filling star while HMXBs are mostly fed directly from the winds of a
$\gtrsim$10 M\sun~companion. They display distinct Galactic
distributions associated with Population I and Population II objects,
with HMXBs lying along the Galactic plane and LMXBs clustering towards
the Galactic bulge and in globular clusters \cite{joss84}
(Fig. \ref{fig:1} ).  A handful of X-ray binaries with $\approx1-3$
M\sun~Roche-lobe filling companions are sometimes referred to as
Intermediate Mass X-ray Binaries (IMXBs).  For a comprehensive review
on X-ray binaries we refer to \cite{charles06}.

The type of X-ray activity observed is determined by (i) the mass
transfer rate from the donor, (ii) the magnetic field of the compact
star, and (iii) the X-ray heating of the accretion disc by the
accretion luminosity. The interplay between these three quantities
explains why black hole remnants are mostly found in transient LMXBs,
neutron stars in persistent LMXBs and pulsars in HMXBs.  In recent
years we have seen the discovery of pulsars with millisecond spin
periods in transient LMXBs. These are considered a missing link in
X-ray binary evolution, with neutron stars being spun up by sustained
accretion to become \textit{recycled} pulsars \cite{alpar82,wijnands98}.
A detailed review of X-ray binary evolution with the variety of
evolutionary paths and end products can be found in Chap. 7.13 of this
book.

X-ray binaries present ideal laboratories for examining the physics of
the supernova explosions which formed their compact objects.  The
orbital motion of the stellar companions can be used to weigh the
masses of the supernova remnants. 
Abundance anomalies are often seen in the companion
star atmospheres, demonstrating chemical pollution by the supernova
ejecta. And the spatial motion of the binary possesses information on
the kick velocity imparted by the explosion
itself. These three topics (dynamical masses, kick velocities and
chemical anomalies) and their impact on our understanding of Type Ibc
and Type II supernovae are the scope of this chapter and will be
presented in turn.

\section{Remnant Masses}
\label{sec:2}
The distribution of masses of compact remnants contains the imprints
of the physics of the supernova explosions.  Various aspects, such as
the explosion energy, mass cut, amount of fallback or the explosion
mechanism itself are important for the final remnant mass
distribution. By building the mass spectrum of compact objects in
X-ray binaries we can therefore obtain new insights onto the physics
of core-collapse in Type Ibc and Type II supernovae. In principle,
precise masses can be extracted from eclipsing double-line
spectroscopic binaries using simple geometry and Kepler's laws but
this is not often the case in X-ray binaries.  Note that neutron star
masses in binary radio pulsars are already covered in Chap. 7.4 of
this book and hence are not discussed here.  It should also be noted
that the accretion process responsible for lighting up the X-ray
binaries can in principle change significantly the neutron star mass
in systems where sufficient time is available such as neutron stars in
old LMXBs. On the other hand, the accreted mass is too low to alter
the BH mass significantly and similarly, the neutron star mass in
short lived HMXBs can also not be changed significantly.

\subsection{Pulsar Masses in HMXBs}
\label{subsec:1}

Pulsars in eclipsing binaries present, in principle, the best
prospects for accurate determination of remnant masses. The Doppler
shift of the donor's photospheric lines, combined with timing delays
of the neutron star pulse, allows us to measure the projected orbital
velocities of the two binary components ($K_{\rm opt}$ and $K_{\rm X}$
respectively) thus making them double-lined binaries.  If the pulsar
is eclipsed by the massive donor (a $\approx$40\% chance in incipient
Roche-lobe overflowing systems
)  then the inclination  
angle {\it i} is given by  

\begin{equation}
\sin i  = {\sqrt{1-(R_{\rm opt}/a)^2}\over{\cos \theta}} 
\end{equation} 

\noindent
where $\theta$ is the eclipse half-angle, $a$ the binary separation and $R_{\rm opt}$ the stellar radius. The latter can be approximated by some fraction 
$\beta\leq1$ of the effective Roche lobe radius $R_{\rm Lopt}$, also known as the stellar "filling factor",  
while $R_{\rm Lopt}/a$ is purely a function of  the binary mass ratio $Q=M_{\rm X}/M_{\rm opt}=K_{\rm opt}/K_{\rm X}$ and the degree of stellar 
synchronization $\Omega$ (usually $\Omega\approx$1). 
The stellar masses can then be solved from the mass function equations 

\begin{equation}
M_{\rm opt}={K_{\rm X}^{3} P(1-e^2)^{3/2}\over{2 \pi G \sin^3 i}}  \left(1+Q\right)^2
\end{equation}

\begin{equation}
M_{\rm X}={K_{\rm opt}^{3} P(1-e^2)^{3/2}\over{2 \pi G \sin^3 i}}  \left(1+{1\over{Q}}\right)^2
\end{equation} 

\noindent
where $P$ stands for the binary period and $e$ the orbital eccentricity.  
This method has produced nine pulsar masses with relatively high
precision which we list in Table \ref{tab:1}. The major source of
uncertainty arises from the combined effect of variable stellar wind,
tidal pulsations and X-ray irradiation which distort the absorption
profiles and hence the radial velocity curve of the optical companion
(e.g. \cite{quaintrell03,reynolds97}).  Although not a pulsar, we have
also included in this section a remnant mass determination for the
eclipsing HMXB 4U 1700-37. With a mass significantly higher than the
nine HMXBs pulsars, the nature of the compact object in this system is
unclear and a low-mass black hole cannot be dismissed. In any case,
the quoted mass should be regarded as somewhat less secure because it
rests upon the spectroscopic mass of the optical companion (see
\cite{clark02} for details).

Interestingly, the largest known population of pulsar X-ray binaries (over 100) belong to the subclass of 
Be/X-ray binaries \cite{reig11}. 
These systems generally have very wide and eccentric orbits, with the neutron star accreting material from the circumstellar Be disc during periastron passages or through episodic 
disc instability events. Unfortunately, the scarcity of eclipsing systems and the very long orbital periods makes reliable mass determination in Be/X-ray binaries extremely difficult.

\begin{table}
\caption{Pulsar and Neutron Star (NS) masses in X-ray Binaries$^{\dagger}$ }
\label{tab:1}      
\begin{tabular}{p{3cm}p{2.4cm}p{2cm}p{2cm}p{2cm}}
\hline\noalign{\smallskip}
Object & X-ray Binary Class & Remnant &  Mass (M\sun) & References \\
\noalign{\smallskip}\svhline\noalign{\smallskip}
OAO 1657-415 & HMXB/persistent & X-ray pulsar & 1.42$\pm$0.26  & \cite{mason12} \\ 
SAX 18027-2016 & ,, & ,, & 1.2-1.9 & \cite{mason11} \\ 
EXO 1722-363 &  ,,  &  ,,  & 1.55$\pm$0.45 & \cite{mason10} \\ 
4U 1538-52 & ,, & ,, & 1.00$\pm$0.10 & \\ 
SMC X-1 &  ,,  &  ,,  & 1.04$\pm$0.09 & \\ 
Vel X-1 & ,, & ,, & 1.77$\pm$0.08 & \\ 
LMC X-4 & ,, & ,, & 1.29$\pm$0.05 & \\ 
Cen X-3 & ,, & ,, & 1.49$\pm$0.08 & \\ 
4U 1700-37 & ,, & ? & 2.44$\pm$0.27 & \cite{clark02} \\ 
Her X-1 & IMXB/persistent & X-ray pulsar & 1.07$\pm$0.36 & \\ 
\noalign{\smallskip}\svhline\noalign{\smallskip}
Cyg X-2 & LMXB/persistent & NS &  1.71$\pm$0.21 & \cite{casares10} \\ 
V395 Car & ,, & ,, & 1.44$\pm$0.10 & \\ 
Sco X-1 & ,, & ,, & $<$1.73 & \cite{mata15} \\ 
XTE J2123-058 & LMXB/transient & ,, & 1.46$^{+0.30}_{-0.39}$ & \cite{tomsick02} \\ 
Cen X-4 & ,, & ,, & 1.94$^{+0.37}_{-0.85}$ & \cite{shahbaz14} \\ 
4U 1822-371 & ,, & X-ray pulsar & 1.52-1.85 & \cite{munoz08} \\ 
XTE J1814-338 & ,, & msec ,, & 2.0$^{+0.7}_{-0.5}$ & \cite{wang17} \\ 
SAX J1808.4-3658 & ,, & ,, ,, & $<$1.4 & \\ 
HETE 1900.1-2455 & ,, & ,, ,, & $<$2.4 & \\ 

\noalign{\smallskip}\hline\noalign{\smallskip}
\end{tabular}
$^{\dagger}$ Adopted from \cite{zhang11} and \cite{rawls11}, unless otherwise stated in the reference column.
\end{table}

\subsection{Neutron Star Masses in LMXBs}
\label{subsec:2}
Neutron stars in LMXBs do not usually pulse (with 4U 1822-371 and a handful of millisecond pulsars as the only exceptions) and their radial velocity curves are thus not available. 
Only the mass function of the compact star is attainable through the radial velocity curve of the optical companion (eq. 3). In these cases it is still possible to derive reliable 
masses by exploiting the fact that the low-mass donor star overflows its Roche lobe and is synchronized 
in a circular orbit (which in turn implies $\Omega=$1, $e=1$). This is reasonable assumption given the long lifetimes and short circularization timescales expected in LMXBs 
\cite{witte01}.  
On this basis, the broadening of the donor absorption lines $V \sin i$ depends on binary mass ratio $q=Q^{-1}$ according to \cite{wade88} 

\begin{equation}
V \sin i/K_{\rm opt}\simeq 0.462~q^{(1/3)} (1+q)^{(2/3)}
\end{equation}

\noindent
while its orbital light curve (governed by tidal distortions) correlates with inclination. 
Therefore, the detection of the faint donor star in LMXBs ensures a full dynamical solution which makes this technique feasible for transient LMXBs in quiescence (i.e. when accretion is halted and X-ray emission is weak) or persistent LMXBs with long orbital periods and thus luminous companion stars. 

In the case of persistent LMXBs with short periods $\lesssim$1 d the companion star is totally overwhelmed by the accretion luminosity. However, some constraints on stellar masses can still be derived through the Bowen technique  which employs fluorescence lines excited on the X-ray heated face of the donor star \cite{steeghs02}. 
The radial velocity curves of the Bowen lines are biased because they arise 
from the irradiated face of the star instead of its center of mass. Therefore, a {\it K-correction} 
needs to be applied in order to recover the true velocity semi-amplitude $K_{\rm opt}$.  
The {\it K-correction} parametrizes the displacement of the center of light with 
respect to the donor's center of mass through the mass ratio and disc 
flaring angle $\alpha$, with the latter dictating the size of the disc shadow 
projected over the irradiated donor~\cite{munoz05}.  
Extra information on $q$ and $\alpha$ is thus required to measure the real 
$K_{\rm opt}$ . Further limits to neutron star masses can be set if the binary 
inclination is well constrained through eclipses (e.g. 4U 1822-371). 
It is interesting to note that the Bowen technique, despite its limitations, has enabled the first dynamical 
constraints in persistent LMXBs since their discovery 50 years ago. 
The best neutron star masses in LMXBs obtained by means of these techniques are also listed in Table \ref{tab:1}.

\subsection{Black Hole Masses}
\label{subsec:3}

The great majority of accreting black holes are found in transient
LMXBs/IMXBs, as proved by dynamical studies.  Relatively precise
masses have been measured in quiescence through exploiting the
photometric and spectroscopic detection of the companion star.  These
are listed in Table \ref{tab:2}. Only in the case of GX 339-4 is the
constraint on the black hole mass aided by the Bowen 
technique.  In some cases uncertainties are quite large owing to
possible systematics affecting the determination of the binary
inclination angle.  In others, only a robust lower limit to the black
hole mass is secured by the spectroscopic mass function combined with
the absence of X-ray eclipses.
\begin{table}
\caption{Black Hole Masses in X-ray Binaries$^{\dagger}$}
\label{tab:2}       
%
%
\begin{tabular}{p{3cm}p{3cm}p{2cm}p{2cm}}
\hline\noalign{\smallskip}
Object & X-ray Binary Class & Mass (M\sun) & References \\
\noalign{\smallskip}\svhline\noalign{\smallskip}
GRS 1915+105 & LMXB/transient &  12.4$^{+2.0}_{-1.8}$  & \cite{reid14} \\ 
V404 Cyg        &  ,, &  9.0$^{+0.2}_{-0.6}$  &  \\ 
BW Cir               & ,, & $>$7.0  &  \\ 
GX 339-4 & ,, & $>$6.0  &  \\ 
XTE J1550-564   & ,, & 7.8$-$15.6 &  \\ 
H1705-250       & ,, & 4.9$-$7.9 &  \\ 
GS 1124-684     & ,, & 11.0$^{+2.1}_{-1.4}$ & \cite{wu16} \\ 
GS 2000+250      &  ,, & 5.5$-$8.8  &  \\ 
A0620-00        &  ,, & 6.6$\pm$0.3 &  \\ 
XTE J1650-500   & ,, & 4.0$-$7.3 &  \\ 
GRS 1009-45  & ,, & $>$3.6 & \\ 
XTE J1859+226   & ,, & $>$ 5.42 &  \\ 
GRO J0422+32    & ,, & $>$1.6 &  \\ 
XTE J1118+480    & ,, & 6.9$-$8.2 &  \\ 
XTE J1819.3-2525 & IMXB/transient & 6.4$\pm$0.6 &  \cite{macdonald14} \\ 
GRO J1655-40    & ,, & 5.4$\pm$0.3 &   \\ 
4U 1543-475     & ,, & 2.7$-$7.5 &   \\ 
\noalign{\smallskip}\svhline\noalign{\smallskip}
Cyg X-1 & HMXB/persistent & 14.8$\pm$1.0 &  \\ 
LMC X-1 &  ,, & 10.9$\pm$1.4 &  \\ 
LMC X-3 & ,, & 7.0$\pm$0.6  & \cite{orosz14} \\ 
M33 X-7 & ,, & 15.7$\pm$1.5 &  \\ 
MWC 656 & HMXB/transient (?) & 3.8$-$5.6 & \cite{casares14a} \\ 
\noalign{\smallskip}\hline\noalign{\smallskip}
\end{tabular}
$^{\dagger}$ Adopted from \cite{casares14b} unless otherwise stated in the reference column. 
Lower limits for BW Cir, GRS 1009-45, XTE J1859+226 and GRO J0422+32 are based on the absence of eclipses, 
combined with updated determinations of the mass function and $q$ (when available). The lower limit on GX 339-4 is based on the lack of X-ray eclipses plus constraints provided by the {\it K-correction}.
\end{table}

Five additional black holes have now been established in HMXBs. Although the companion star underfills its   
Roche lobe in these systems, extra information on the stellar radius  is granted by very precise distance determinations ($\lesssim$5\% e.g. through VLBI parallax for Cyg X-1) coupled with the observed apparent brightness and effective temperature. Furthermore, in the case of M33 X-7, eclipses of the X-ray source by the donor star provide additional tight constraints on the inclination which results in one of the largest accurately known black hole masses. 

Note that we have excluded mass measurements for the two extragalactic
HMXBs NGC~300 X--1 and IC~10 X--1. This is because these rely on
radial velocity curves of the HeII $\lambda$4686 wind emission
line and an assumed mass for the Wolf-Rayet star. Different groups
have reported conflicting results which range from canonical neutron
stars to the largest black hole masses measured so far and are hence
unreliable. The table includes MWC 656, the first black hole companion
to a Be star \cite{casares14a}. Here the black hole mass relies on the
spectroscopic mass of the optical companion and the radial velocity
curves of the two stars, extracted from the dynamics of circumstellar
gaseous discs.  A critical review of black hole mass determinations,
including potential systematic effects, is presented in
\cite{casares14b}.

\begin{figure}[h] \hbox{\hspace{-0.1cm}
\vspace{-2.5cm}
\includegraphics[angle=0,width=10cm]{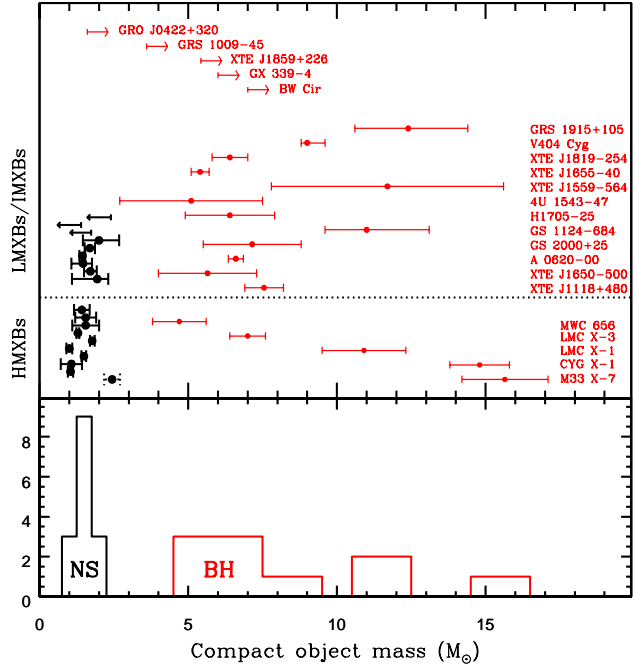}
\hspace{1.0cm}} \caption{
Top: compact remnant masses measured in X-ray binaries. Neutron stars and black holes are 
indicated in black and red colours, respectively. 4U 1700-37 is plotted 
in dotted-style line because the nature of the compact star is uncertain. 
The horizontal dotted line divides LXMBs/IMXBs from HMXBs. 
Bottom: Observed distribution of neutron stars and black hole masses.}
\label{fig:2}       
\end{figure}

\subsection{The Mass Spectrum. Implications for Supernovae Models}
\label{subsec:4}

Figure \ref{fig:2} presents the observed masses of neutron star and
black hole remnants in X-ray binaries, as in Tables \ref{tab:1} and
\ref{tab:2}.  Note that we have here excluded neutron star masses in
radio pulsars and binary millisecond pulsars as these are covered by
Chap. 7.4.  The bottom panel displays the number distributions of
neutron stars (black) and black holes (red) masses, excluding
upper/lower limits. Three main features seem to be drawn from the
plot, namely (1) neutron star masses tend to be larger in LMXBs/IMXBs
(mass average of 1.54$\pm$0.16 M\sun) than in HMXBs (1.34$\pm$0.26
M\sun), (2) a dearth of remnants or gap appears between $\sim$2-5
M\sun, and (3) the most massive black holes ($\sim$15 M\sun) are found
in HMXBs.

Feature (1), although tentative, could be a manifestation of the
pulsar recycling scenario. The difference in neutron star masses, if
confirmed, would stem from different binary evolution histories, with
neutron stars in LMXBs having experienced significant accretion over
extended periods of time.
This interpretation would be further supported by indications that pulsar
mass decreases with spin period \cite{zhang11}.  Neutron stars in
HMXBs are little modified by accretion and, thus, their masses are
expected to lie closer to their birth values. And indeed, both the mean and
dispersion of the neutron star mass distribution in HMXBs are found to
agree with theoretical expectations of core-collapse supernovae
\cite{ozel12}.  Constraints on neutron star forming supernovae do seem
to be provided by two distinct populations of X-ray pulsars in
Be/X-ray binaries; short $P_{\rm spin}$ pulsars with short orbital
periods and low eccentricities would be produced by electron-capture
supernovae while long $P_{\rm spin}$ pulsars with long orbital periods
and high eccentricities in iron-core-collapse supernovae
\cite{knigge11}. The former pulsars are naturally expected to be less
massive ($\lesssim$1.3 M\sun) but, unfortunately, this cannot yet be
tested because of the lack of precise neutron star mass determinations
in Be/X-ray binaries.

\begin{figure}[h] \hbox{\hspace{-0.1cm}
\vspace{-2.5cm}
\includegraphics[angle=0,width=10cm]{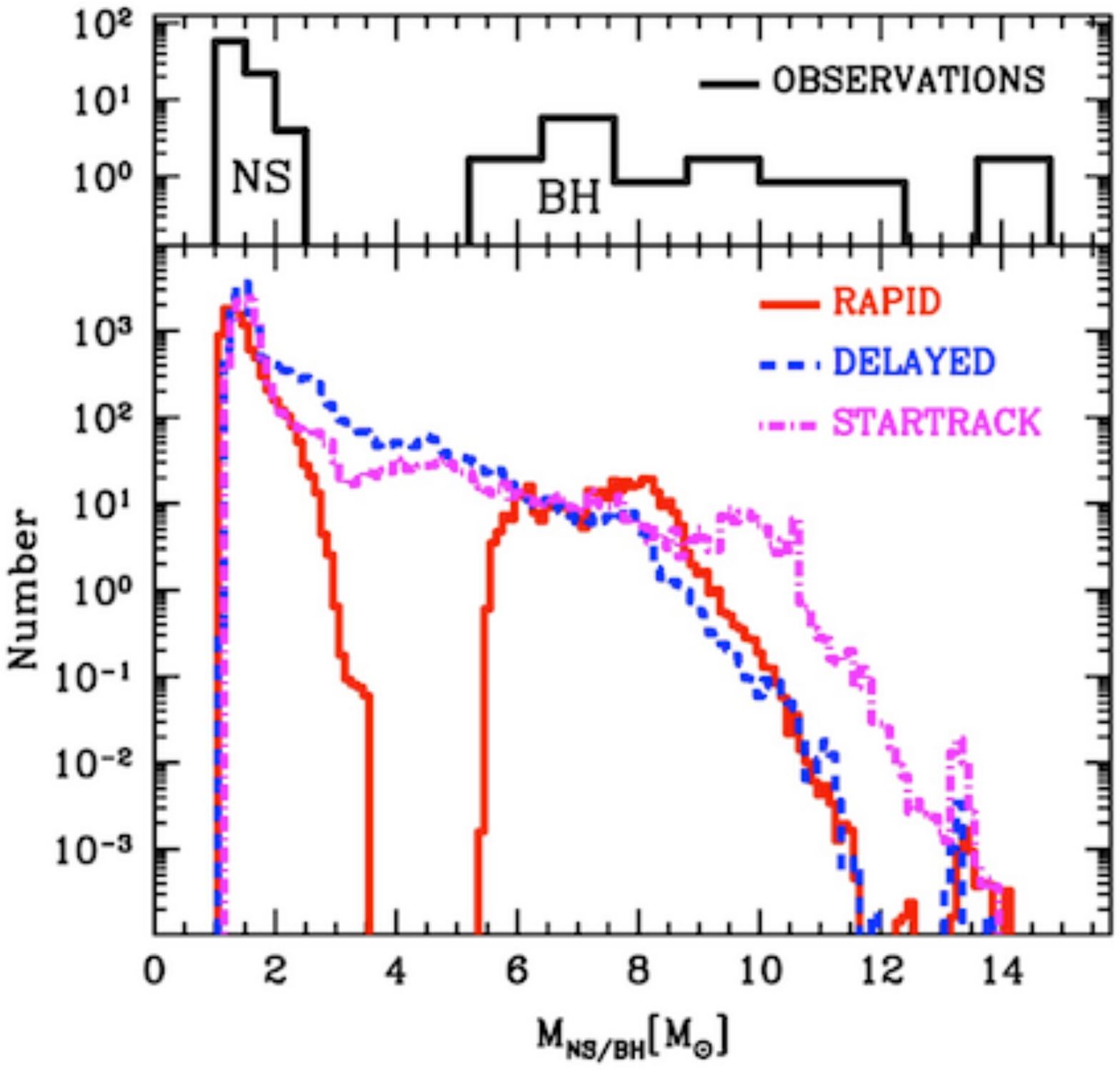}
\hspace{1.0cm}} \caption{
Observed mass distribution of compact objects in X-ray binaries (top), compared to theoretical distributions computed  for different supernova models (bottom). 
The mass gap can be reproduced only if turbulent instabilities grow rapidly. Slow growing instabilities lead to significant fallback that would fill the mass gap. From 
\cite{belczynski12}. }
\label{fig:3}       
\end{figure}

Feature (2) is a statiscally robust property of the mass spectrum (see \cite{farr11} and references therein). 
The lack of compact remnants between $\sim$2-5 M\sun~contrasts with numerical simulations of supernova 
explosions by \cite{fryer01} that lead to continuous distributions and typical exponential decays. These simulations, however, are based on single star populations 
with a heuristic treatment of binarity through Wolf-Rayet winds following 
common envelope evolution. In order to accommodate the evidence 
of a mass gap, a discontinuous 
dependence of explosion energy with progenitor mass seems unavoidable. 
In this context, it has been proposed that convection (Rayleigh-Taylor)  
instabilities, growing within 200 ms after core bounce, can successfully revive the supernova shock 
and trigger the explosion, thereby causing the 
gap (see Fig. \ref{fig:3}, but see also \cite{ugliano12} for a different interpretation based on 
neutrino-driven explosion models).  
Alternatively, a gap 
can be produced if red super-giant stars of $\approx$17-25 M\sun~suffer a failed supernova explosion, 
leaving a remnant with the mass of the He core while ejecting the weakly bound H envelope \cite{kochanek14}. 
This interpretation appears attractive because in turn it provides an explanation for 
the deficit of high-mass progenitors seen in pre-explosion images of Type IIp supernovae \cite{smartt09}. 
On the other hand, it fails to account for the peculiar abundance of $\alpha$-elements detected in the companion stars which 
demands significant contamination from supernova ejecta (see Sect. \ref{sec:4}).  
It is also unclear how very wide binaries with such red supergiants can evolve to form 
the compact black hole binaries that we see today.

Feature (3) most likely reflects different binary evolutionary paths, with black holes in LMXBs being limited to $\lesssim$12 M\sun~by 
severe mass loss from the Wolf-Rayet progenitor after the common envelope phase \cite{fryer01}. 
Conversely, black holes in 
HMXBs can grow from more massive stars, especially in low-metallicity environments such as in the case of M33 X-7.  
Furthermore, it is possible that the progenitor star evolves through the He burning phase still embedded in the H envelope 
(case C mass transfer), thus suffering less wind mass-loss \cite{brown01}. 
However, it should be noted that some aspects of binary and massive stellar evolution (e.g. radial expansion, wind mass-loss rates, efficiency of 
common envelope ejection) are still quite uncertain, which certainly limits our understanding of the formation of X-ray binaries. 

At this point, the impact of systematic uncertainties in the determination of binary inclination angles should also be stressed, 
as exemplified by the large dispersion of values 
reported by independent groups on individual systems (see \cite{casares14b}). Ignoring the role of systematics can lead to overestimated black hole masses 
and hence a bias in the observed distribution \cite{kreidberg12}. In addition, the sample of X-ray binaries with dynamical mass determinations is almost certainly prone to 
complex selection effects (both evolutionary and observational) not completely understood. For instance, it may be possible that low-mass black holes become unbound by the 
supernova explosion or are hidden in very faint (but persistent) X-ray binaries. The system MWC 656 might be itself a member of a hidden population of very faint low-mass black holes.    
More observational work is required to enlarge the sample of secure black hole masses before the observed 
distribution can be definitively used to illuminate the properties of the supernova engine. 

\section{Kick Velocities}
\label{sec:3}

For this Section we focus on stellar--mass black holes in X--ray binaries
rather than neutron star X--ray binary systems as their space velocities are more
readily measured using pulsar timing information \cite{gonzalez_pj_11}.
Both the black hole mass and its spin as well as the black hole's
velocity through space are almost exclusively set at the instant of formation. The
black hole spin and mass do not change appreciably during its subsequent 
evolution. Even the accretion of material that allows us to discover
these sources as black hole X--ray binaries 
when they outburst does not change these parameters. The reason is that in
order to affect the black-hole spin or mass significantly one needs to
accrete of order the black hole mass. Due to the Eddington limit and
the limited amount of mass available from a stellar--mass donor star
this is not possible. 
Encounters with giant molecular clouds, and
spiral density waves result in changes in the gravitational potential
\cite{1977A&A....60..263W} and these {\it can} change the space
velocity of stellar--mass black holes in X--ray binaries as they
travel through our Galaxy in the Gyrs after their formation.  The
magnitude of velocity changes depends on the age of the system and can
at most amount to up to 40 km s$^{-1}$ as determined for late type
(single) stars from Hypparcos data by \cite{1998MNRAS.298..387D}.
Thus, with the caveat on space velocities less than $\sim40$ km
s$^{-1}$, these parameters are a prior for the black hole formation
mechanism, providing input on the black hole formation and supernova
models.

\subsection{Black hole formation and its link to space velocity}
Models for black hole formation distinguish two formation scenarios
(\cite{fryer01}): 
either direct or delayed collapse where
in the latter a neutron star is formed first which moments later
collapses into a black hole. The more massive progenitor stars could
collapse directly to a black hole without producing a strong
supernova. The latter scenario is in line with the absence of
supernovae from red--supergiants more massive than $\sim 18-20$
M$_\odot$ (\cite{2015PASA...32...16S}). 
Instead of a supernova, the
event marking the birth of a black hole could be a faint,
short--duration (3--10 days) blue transient that may form from
direct-collapse red-supergiant progenitors as the shock caused by
the response of the stellar envelope to neutrino emission in the
collapsing core equivalent to a few times 0.1 M$_\odot$ of rest--mass
energy breaks out of the star (\cite{2013ApJ...768L..14P}). Others
propose that also a red transient with a longer time scale will ensue,
which should be of the order of a year
(\cite{2011ApJ...730...70O}; \cite{2013ApJ...769..109L}). Candidate
failed supernovae of a red--supergiant have been reported; one in
NGC~6946 (\cite{2015MNRAS.450.3289G}) and one in NGC~3021
(\cite{2015MNRAS.453.2885R}). Given the brightness of red--supergiants
in the near--infrared, a dedicated near--infrared variability study of
nearby galaxies could have a good chance of finding such events,
even though some Mira and R Coronae Borealis variables may have similar
characteristics (although in the failed supernova \& black hole
formation case the transient should fade away indefinitely).

The main uncertainties in the theoretical supernova and (binary)
massive star evolution calculations stem from uncertainties in the
supernova explosion mechanism for stars with progenitor masses in the
range between 11 and approximately 30 M$_\odot$ and also from
uncertainties in the amount of mass lost during the evolution for
stars with progenitor masses above approximately 30
M$_\odot$. Neutrino driven supernova models
(e.g.~\cite{ugliano12}) seem compatible with the observed
black hole mass distribution reported in 
\cite{2010ApJ...725.1918O} and \cite{farr11}
with an apparent lack of low--mass black
holes in the mass range of 2--5 M$_\odot$ (see Section~\ref{sec:2}).

The direct collapse and the delayed proto-neutron star collapse
models might be responsible for the formation of different mass black
holes, and also different space velocities. Direct collapse produces a
black hole without much of a kick, whereas the delayed supernova
models produce relatively low--mass black holes that receive a larger
kick and thus space velocity.

\subsection{The black hole space velocity}

The difference 
between the velocity of a (black hole)
system and that expected for its local standard of rest is called the
space velocity. Occasionally, this velocity is called ``peculiar
velocityÕÕ,  but this term is also often used to indicate the velocity
difference between the Hubble flow and the velocity of galaxies.  Hence
to avoid confusion we use ``space velocityÕÕ for the black hole binary
systems.

It has been inferred from the velocity distribution of neutron stars
observable as single radio pulsars that they receive a kick at birth
due to asymmetries in the supernova explosion 
\cite{1994Natur.369..127L}. This we call a natal kick. Black holes
forming from fall-back onto a proto--neutron star should then also
get such a natal kick, resulting potentially in a significant space
velocity. The magnitude would depend on whether the natal kick
momentum for a neutron star and proto-neutron star that collapses to a
black hole is the same or not. In addition to such a natal kick, in
any supernova explosion in a binary where mass is lost from the binary
system, a kick should be imparted on the system irrespective of the
type of compact object formed during the supernova
\cite{1961BAN....15..265B}. 
The binary will be disrupted if more than half the
total binary mass is ejected in the supernova.
Therefore,
this sets an upper limit to the amplitude of such a so called 
Blaauw kick \cite{1999A&A...352L..87N}. Information that can help
differentiating between a Blaauw and a natal kick is that a Blaauw
kick is directed in the binary plane (unless the supernova mass loss
is not symmetric) whereas a natal kick may not be restricted to that
plane. Given that large natal kick velocities yield a higher
probability to unbind the binary system, a population synthesis model
explaining the observed black hole X--ray binary mass and space
velocity distribution has to be used to correct for this bias
(e.g.~\cite{belczynski12}).

Evidence for velocity kicks in neutron star as well as black hole
low--mass X--ray binaries, comes from the fact that they have, on
average, a large distance to the plane of the Galaxy
(\cite{1995ApJ...447L..33V}; \cite{2004MNRAS.354..355J};
\cite{2015MNRAS.453.3341R}). Evidence for kicks in the formation of
LMXBs also comes from their observed distribution in early type
galaxies. The LMXBs extend further than the stellar light consistent
with a population of kicked LMXBs \cite{zhang13}. In contrast, given
that LMXBs are also found in the Large Magellanic Cloud, and perhaps
dwarf galaxies (cf.~\cite{2005MNRAS.364L..61M}) a fraction must
receive a small kick velocity upon formation otherwise they would not
be found in those systems given their low escape
velocities. Similarly, the recent evidence for the presence of black
holes in globular clusters (e.g.~\cite{2012Natur.490...71S};
\cite{2013ApJ...777...69C}) implies a population of low natal kick
black holes.

\subsection{Measuring the space velocity}
For individual sources the space velocity can be determined as
follows: the optical or near--infrared spectroscopic observations
providing the mass function via the measurement of the radial velocity
amplitude (see Section~\ref{sec:2}) will also provide the systemic
radial velocity if, in the cross--correlation of the source and
template spectra, the template is a radial velocity standard  
or a model atmosphere.  In order to calculate the
space velocity of the X--ray binary its systemic radial velocity has
to be combined with a proper motion and distance measurement.

Ideally, as it depends on geometry alone, the distance is measured
through a (radio) parallax measurement. However, in practice parallax
measurements have so far only been possible for three black hole
X--ray binaries: V404~Cyg \cite{2009ApJ...706L.230M}; Cyg~X--1
\cite{2011ApJ...742...83R} and GRS~1915$+$105
\cite{reid14}. The main reasons why a parallax
measurement has been possible for these three systems is that V404~Cyg
is bright in quiescence when compared to the other systems
(cf.~\cite{2011ApJ...739L..18M}), and the latter two black holes are
virtually always actively accreting matter, which makes that the time
baseline necessary for a parallax measurement is long enough. However,
while always accreting, the last two sources are not always in the
accretion state that allows a compact radio jet to be formed, which
still made the detection of the parallax signal difficult (see the
discussion in \cite{reid14}). For the vast majority of
black hole X--ray transients a parallax measurement has not been
possible. A major obstacle in the measurements has been the short
duration of outbursts. The months--long outburst coupled with the
aforementioned state changes do not allow for the measurements
necessary to detect the often small parallax signals. The best future
hope for parallax signal detection comes from a few sources that show
recurrent outbursts such as GX~339--4.

The second--best distance determination comes from what is often
called a photometric parallax. Here, one compares the apparent
magnitude of the companion (=mass donor) star with its absolute
magnitude to determine the distance. However, the effect of
interstellar extinction influences the apparent magnitude, it causes
the star to appear fainter and redder and hence one would put it
further away than it really is if these effects are not corrected
for. Furthermore, more often than not, residual light is produced by
the accretion disc, making the system appear brighter. The disc
contribution can be determined from optical spectroscopic observations
but it is not constant in time \cite{2008ApJ...673L.159C},
complicating the determination of the correction factor. The fact that
the star is loosing mass influences the radius of the star and as such
it is not the same as that of a single star of the same spectral type
and luminosity class. Hence, its radius has to be determined from the
data. Finally, in accreting black hole X--ray binary systems such as
under consideration here, the star is very likely to be in forced
co--rotation with the orbit making it rotate around its axis fast
which may influence the absolute magnitude of the star. Studies of
rapidly rotating early type stars found that fast rotation causes an
increase in absolute magnitude of stars with spectral type later than
B5 (the stars are intrinsically less luminous than the non--rotating
stars of the same spectral type by typically several tenths of a
magnitude \cite{1977ApJS...34...41C}).

Additionally, it is known that limb and gravity darkening effects
change the equivalent widths of photospheric stellar absorption lines
(e.g.~\cite{1929MNRAS..89..222S}; \cite{1977ApJS...34...41C}), which
could lead to a slightly erroneous spectral type being obtained from
the data. Generally, the lines in the spectrum resemble the lines of a
later spectral type. Furthermore, the limb and gravity darkening are
different in the distorted Roche lobe filling mass--donor stars
compared to single stars. Overall, distances determined via a
photometric parallax can easily be off by several tens of per
cent. See \cite{2004MNRAS.354..355J} for further discussion on these
issues.

\begin{figure}[h] \hbox{\hspace{-0.1cm}
\includegraphics[angle=0,width=10cm]{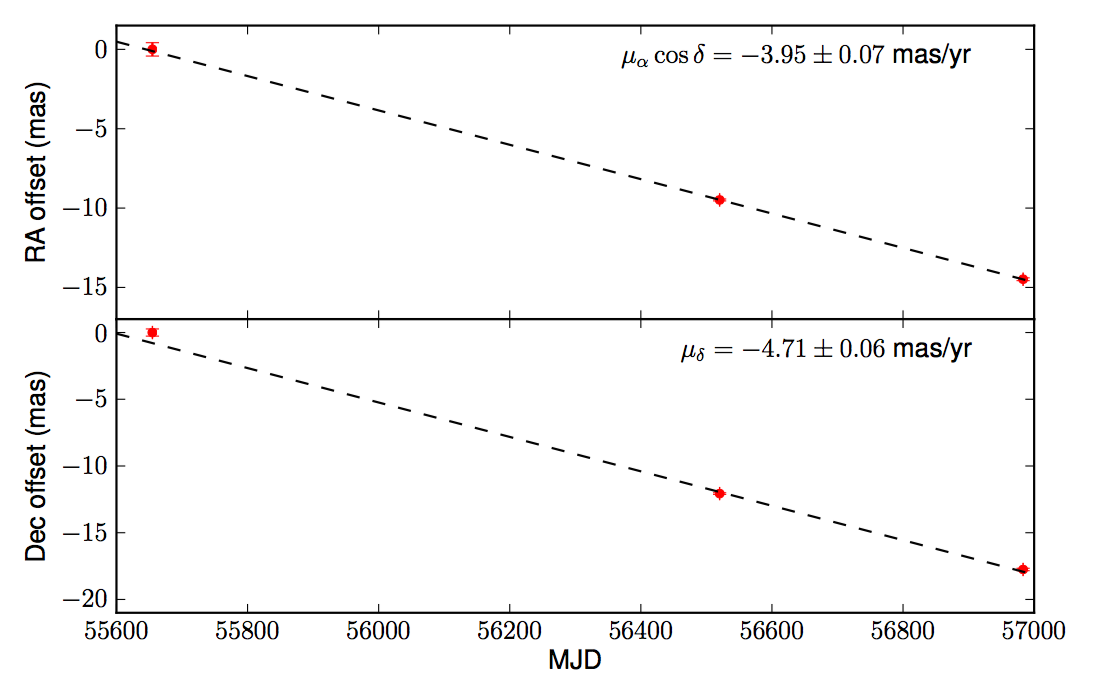}
\hspace{1.0cm}} \caption{The proper motion fit to Long Baseline Array
radio observations of GX~339--4. Whereas the data is sparce, mostly
due to the outburst duty cycle of this source, the proper motion
signal is significantly detected (figure courtesy James Miller--Jones;
Miller--Jones et al.~in prep.)}\label{339-4}\end{figure}

In contrast with the parallax signal, the accuracy with which proper
motions can be detected is higher as the proper motion signal adds
over time. The proper motion has been measured using radio very--long
baseline interferometry (VLBI) of eight black hole X--ray binaries
(see Table~\ref{tab1:sec3}). Recent measurements show the proper
motion of the recurrent transient black hole GX~339--4 (Miller-Jones
et al.~in prep., see Fig. \ref{339-4}). For this source we still need to determine the
spectral type of the mass donor star and the accretion disc
contribution to the total light in quiescence in order to estimate the
distance to the system more accurately using a photometric parallax
than what is currently possible \cite{2004ApJ...609..317H}.

\begin{table}
\caption{Proper motions of black hole X--ray binaries}
\label{tab1:sec3}      
\begin{tabular}{p{2.5cm}p{2.4cm}p{2cm}p{4.9cm}}
\hline\noalign{\smallskip}
Source name & $\mu_\alpha\cos \delta$ (mas yr$^{-1}$) & $\mu_\delta$
                                                        mas yr$^{-1}$ & References \\
\noalign{\smallskip}\svhline\noalign{\smallskip}
XTE~J1118$+$480 & -16.8$\pm$1.6  & -7.4$\pm1.6$ & \cite{2001Natur.413..139M} \\
GRO~J1655--40 & -3.3$\pm$0.5 & -4.0$\pm$0.4& \cite{2002A&A...395..595M}\\
GX~339--4 & -3.95$\pm$0.07 & -4.71$\pm$0.06 & $^{1}$\\
Swift~J1753.5--0127 & 1.5$\pm$0.4 & -3.0$\pm$0.4 & $^{1}$\\
MAXI~J1836-194 & -2.3$\pm$0.6 & -6.1$\pm$1.0 & \cite{2015MNRAS.450.1745R}\\
GRS~1915$+$105  & -2.86$\pm$0.07 & -6.20$\pm$0.09 & \cite{2007ApJ...668..430D}\\
Cyg~X--1 & -3.78$\pm$0.06 & -6.40$\pm$0.12 & \cite{reid14}\\
V404~Cyg & -5.04$\pm$0.22 & -7.64$\pm$0.03 & \cite{2009ApJ...706L.230M}\\
\noalign{\smallskip}\hline\noalign{\smallskip}
\end{tabular}
$^{1}$Miller--Jones et al.~in prep.\\
\end{table}

\begin{figure}[h] \hbox{\hspace{-0.1cm}
\includegraphics[angle=0,width=10cm]{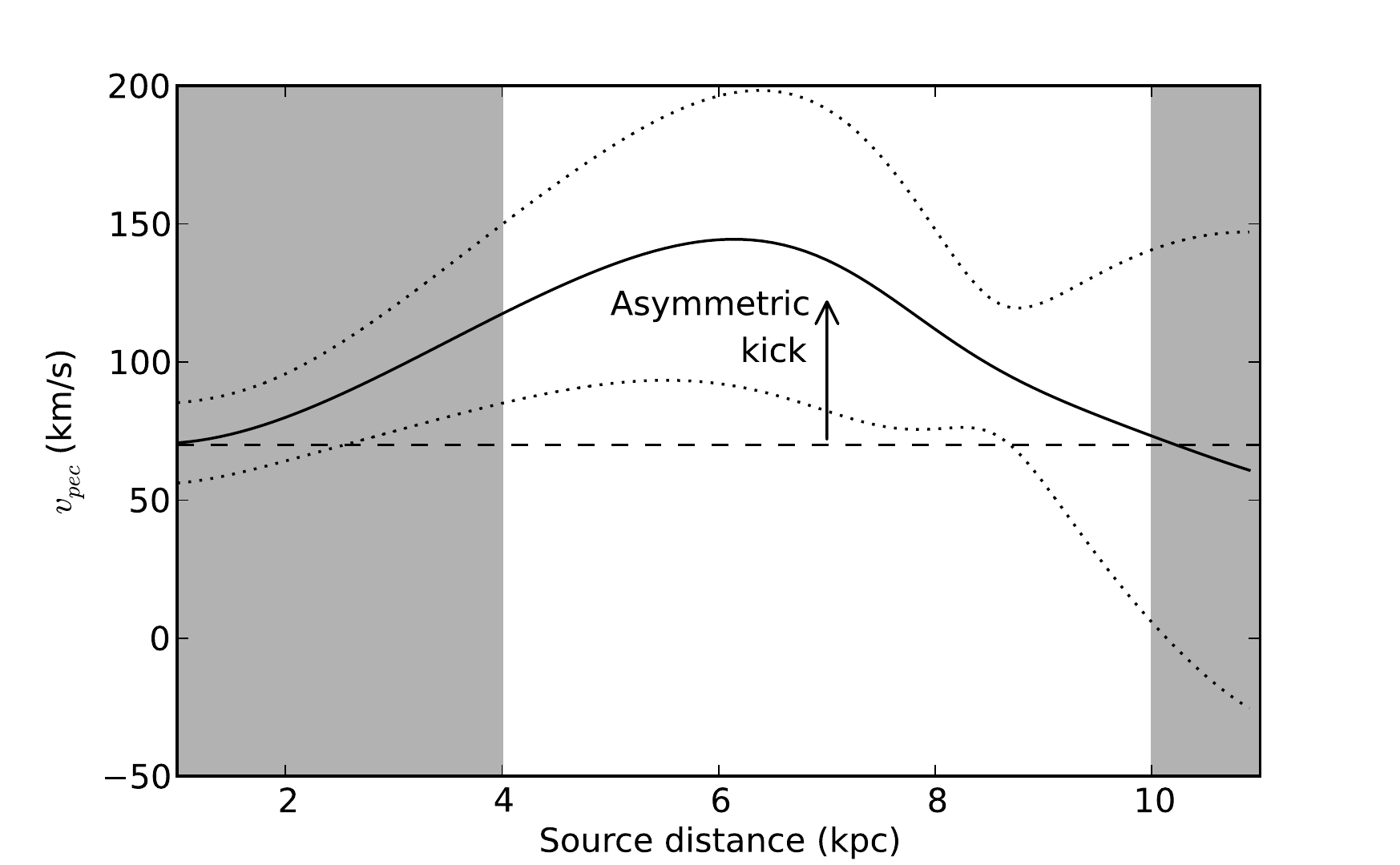}
\hspace{1.0cm}} \caption{ The asymmetric natal kick exerted on the
black hole upon formation on top of the maximum Blaauw kick possible
without disrupting the binary (dashed line;
\cite{1999A&A...352L..87N}) required to explain the observed proper
motion for the black hole candidate source MAXI~J1836--194
(\cite{2015MNRAS.450.1745R}). The dotted lines indicate the
measurement uncertainties and the white area is the range in distances
that is allowed for this particular source. The observables that still
need to be measured to determine the space velocity are the systemic
radial velocity and the source distance. A model for the Galactic
rotation is used \cite{2012PASJ...64..136H}. 
} \label{spacevelo}\end{figure}

In some cases the location on the sky together with a proper motion
measurement can already put strong constraints on the presence of
natal kicks.  For instance, for the black hole candidate (candidate as
no mass measurement for the source exists currently), MAXI~J1836-194,
evidence from the radio proper motion alone suggests that the source
received a kick at birth (see Figure~\ref{spacevelo}).

Whereas so far all these proper motion measurements have come from
radio VLBI measurements, in the near future the Gaia satellite may
also provide proper motion measurements of X--ray binaries. Gaia has
been launched in December 19, 2013 and scans the whole sky (including
the Galactic plane) at high spatial resolution with accurate
photometry down to G=20.3 and tens of micro--arcsecond astrometry.
Gaia consists of two telescopes aligned in a plane with an angle of
106.5 degrees in between. As Gaia scans the sky it will make many
visits of the same region. For most of the sky the number
of visits is $\approx$70 over a 5--year mission lifetime (and an
extension of the mission is possible improving primarily the proper
motion measurements). The multiple visits of the same parts of the sky
will allow the detection of proper motion and parallax signals if the
sources are bright enough. Gaia's $G$--band is effectively a white
light band where the bandpass is set by the efficiency of the
telescope $+$ CCD.

Proper motion and parallax measurements of several of the black hole
X--ray binaries mentioned in Table~\ref{tab1:sec3} will be improved.
In addition these parameters will be measured for a few sources for
which these measurements do not exist currently. An example of the
latter is 1A~0620--00. However, the intrinsic and apparent faintness
of the often low--mass donor stars is a limiting factor for Gaia's
contribution to this field. Many of the black hole X--ray binaries
reside in the plane of the Galaxy and are further away than a few kpc,
making reddening significant. The fact that they reside in the plane
probably hints at a low space velocity as sources spend most of the
time at the extremes of their Galactic orbit. In general (non--Gaia)
astrometric measurements in the optical or near--infrared are difficult. 
An obstacle
to the required astrometric accuracy is the need to find astrometric
``anchors'' that tie the frame to the International Celestial
Reference System. Background quasars and active galactic nuclei are
good candidate anchors, but in the plane their apparent magnitude is
reduced also due to the reddening.

Once the position on the sky, the systemic radial velocity, the proper
motion, and the distance to the source are known one can derive the
space velocity using the transformations of \cite{1987AJ.....93..864J}
to calculate Galactic velocities in the heliocentric frame, which can
then be corrected for the space velocity of the Sun with respect to
the local standard of rest (U=10.0$\pm$0.36 km s$^{-1}$,
V=5.25$\pm$0.62 km s$^{-1}$, W=7.17$\pm$0.38 km s$^{-1}$;
\cite{1998MNRAS.298..387D}).  Here, U is defined as positive towards
the Galactic centre, V positive towards $l$=90$^\circ$ and W positive
towards the North Galactic Pole.

\subsection{Black hole X-ray binaries with and without natal kicks}
Having described in some detail how to measure the parameters that are
necessary to determine the space velocity of stellar--mass black holes
in X--ray binaries, we now turn to the measured space velocities. The
space velocities have been determined for five black hole X--ray
binaries only so far. A recent overview of black hole space velocities
is presented in \cite{2014PASA...31...16M} (his table 2). 

Repetto et al. (2015) \cite{2015MNRAS.453.3341R} investigated whether there is
evidence for the presence of a natal besides a Blaauw kick taking the
binary evolution of seven short orbital period systems into
account. The main conclusion is that for at least two systems a natal
kick is necessary. In addition, five systems could be well explained
with a natal kick but virtually no ejection of mass in a supernova,
such as in direct collapse scenarios.

Natal kicks are probably necessary for the black hole X--ray binaries
XTE~J1118$+$480 (\cite{2005ApJ...618..845G}) and GRO~J1655--40
(\cite{2005ApJ...625..324W}), and are likely for V404~Cygni
(\cite{2009MNRAS.394.1440M}). Cygnus~X--1 and GRS~1915$+$105, on the
other hand, were found to have been formed with little or no kick
(\cite{2003Sci...300.1119M}; \cite{2011ApJ...742...83R};
\cite{{reid14}}). 
It is interesting to note that the black holes
in Cygnus~X--1 and GRS~1915$+$105 are among the most massive
stellar--mass black holes known so far in our Galaxy, which could
be interpreted as suggestive evidence for a formation difference
between the more massive and lighter stellar--mass black holes. For
instance, the more massive black holes could form through direct
collapse giving rise to no or only very small space velocities as only
a limited amount of mass is lost from the system (minimizing the
Blaauw kick) and the maximum natal kick impulse imparted on any
proto--neutron star due to asymmetric neutrino emission has an upper
limit which equals the binding energy of the maximum mass
proto--neutron star that can be formed 
\cite{2013MNRAS.434.1355J}. The latter thus would imply that the
more massive the black hole formed, the smaller the space velocity
acquired. However, in a scenario where the more massive black holes
form through a supernova (and not a direct collapse) and when the fall
back of the slowest parts of the supernova ejecta is asymmetric, one
would produce larger natal kicks for more massive black holes
\cite{2013MNRAS.434.1355J}. Space velocity and black hole mass
measurements for more black hole X--ray binaries as well as searches
of failed supernovae are necessary to distinguish between these
scenarios.

\section{Chemical Abundance of Companion Stars}

\label{sec:4}

It is conceivable that the supernova explosion that created a black hole
or a neutron star remnant in X-ray binaries has modified the physical
and chemical characteristics of the secondary star.  
However, uncertainties in the supernova explosion models affect the
predictions of the chemical composition of
ejecta captured by the companion.
Nevertheless, chemical abundance studies of the companion stars in
LMXBs may open a new route to constrain supernova
models. Unfortunately,
high quality spectroscopic observations of LMXBs and their analysis is
a serious challenge. The rotation, possible spots, and magnetic
activity of the companion star, as well as continuum veiling produced
by the residual accretion disc introduce additional
uncertainties. 
In any case, attempts have been made to minimize the
most important sources of uncertainties in the calculation of the
stellar parameters and chemical abundances. Nova Scorpii 1994 was the
first black hole X-ray binary system for which a detailed abundance
study has been carried out \cite{israelian99}.  Striking
overabundance of several $\alpha$-elements (such as O, S, Si) were
discovered and interpreted as a result of a pollution by matter
ejected by the supernova.  

\subsection{Observations, models and spectral synthesis tools }

The chemical analysis of secondary stars in LMXB systems is influenced
by three important factors: veiling from the accretion disc,
rotational broadening and signal-to-noise ratio of the spectra. 
Moderately strong and relatively unblended lines of chemical elements
of interest have to be identified in the high resolution solar flux
atlas. Spectral line data from
the Vienna Atomic Line Database can be used to compute synthetic
spectra for these features employing a grid of local thermodynamic
equilibrium (LTE) models of atmospheres provided by Kurucz (1993,
private communication) \cite{kur93}. These models are interpolated for
given values of effective temperature [$\rm T_{\rm eff}$], surface gravity [$\log$
g], and metallicity [Fe/H]).

A grid of synthetic spectra is generated for these features in terms
of five free parameters three to characterize the star atmospheric
model ([$\rm T_{\rm eff}$], [$\log$ g], and 
[Fe/H]) and two further parameters to take into
account the effect of the accretion disk emission in the stellar
spectrum. This veiling is defined as the ratio of the accretion disk
flux to the stellar continuum flux, $F_{disc}/F_{cont,star}$. It has
been considered as a linear function of wavelength and is thus
characterized by two parameters: veiling at 4500\AA,
$f_{4500} = F^{4500}_{disc} / F^{4500}_{sec}$, and the slope,
$m_{0}$. 

\begin{figure}[h] \hbox{\hspace{-0.1cm}
		\includegraphics[angle=0,width=10cm]{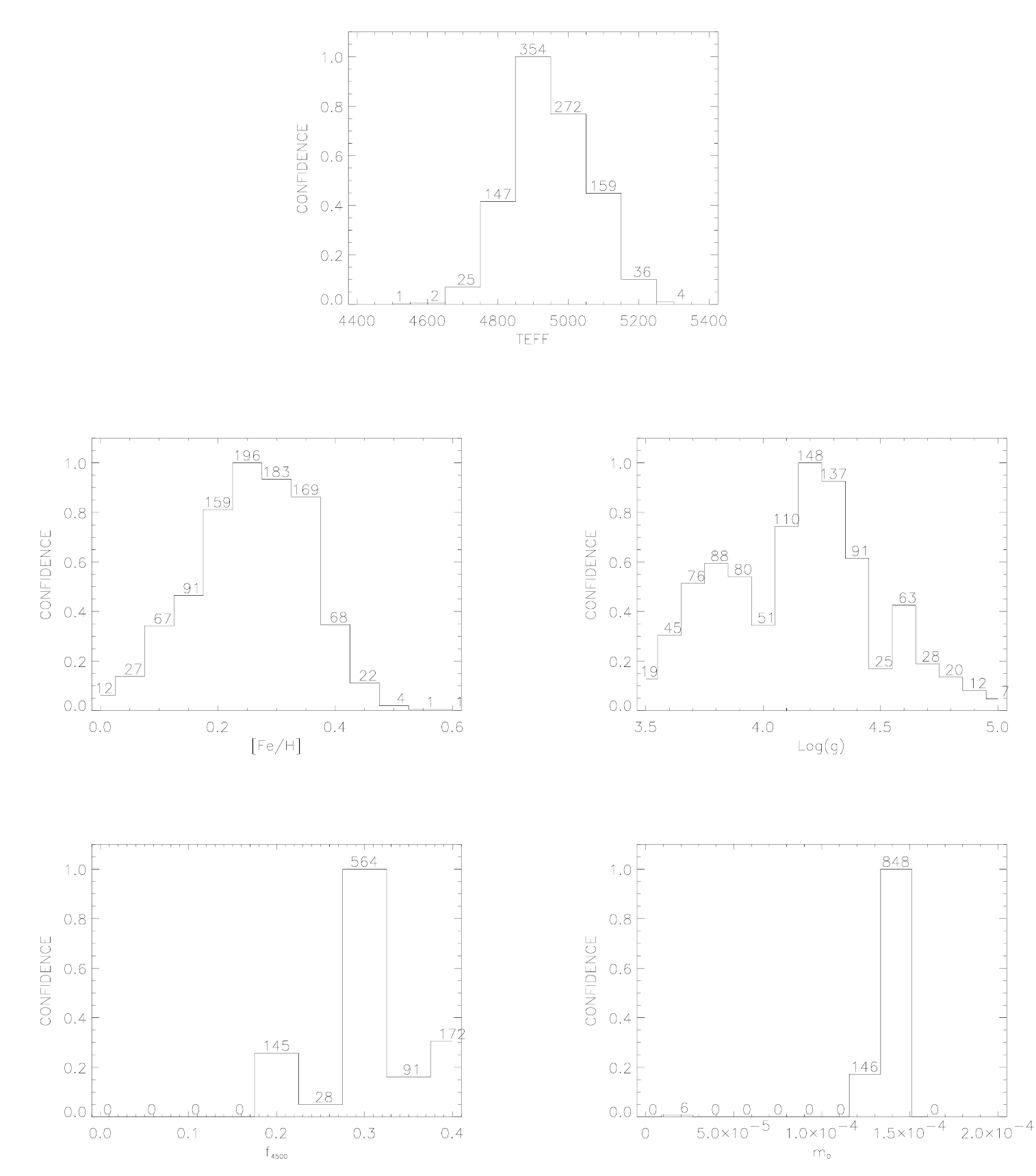}
		\hspace{1.0cm}} \caption{ Distributions obtained for
                each parameter using Monte Carlo simulations. The
                labels at the top of each bin indicate the number of
                simulations consistent with the bin value. The total
                number of simulations was 1000.  Figure from \cite{gonzalez04}.
                }
	\label{A0620}       
\end{figure}

Next, the observed spectra are compared with each synthetic spectrum
in the grid (between 800.000 and 1.5 million spectra) via a $\chi^{2}$
minimization procedure that provides the best model fit. A bootstrap
Monte-Carlo method using 1000 realizations is typically used to define the
1$\sigma$ confidence regions for the five free parameters.
Fig \ref{A0620} shows the distributions obtained for the source
A0620-00. 

\subsection{Stellar abundances}

\begin{figure}[h] \hbox{\hspace{-0.1cm}
		\includegraphics[angle=0,width=10cm]{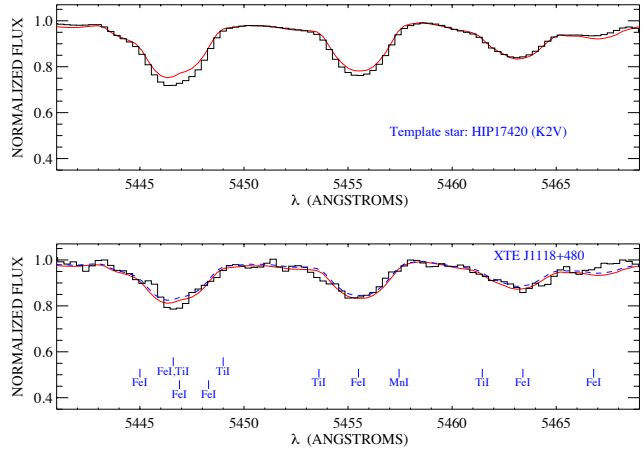}
		\hspace{1.0cm}} \caption{
		Best synthetic spectral fits to the Keck/ESI spectrum of the secondary star in the XTE J1118+480 system
		(bottom panel) and the same for a template star (properly broadened) shown for comparison (top panel). Synthetic
		spectra are computed for solar abundances (dashed line) and best-fit abundances (solid line).  This 
		figure is taken from \cite{gonzalez08b}
	}
	\label{xte}       
\end{figure}

\begin{figure}[h] \hbox{\hspace{-0.1cm}
		\includegraphics[angle=0,width=5.75cm]{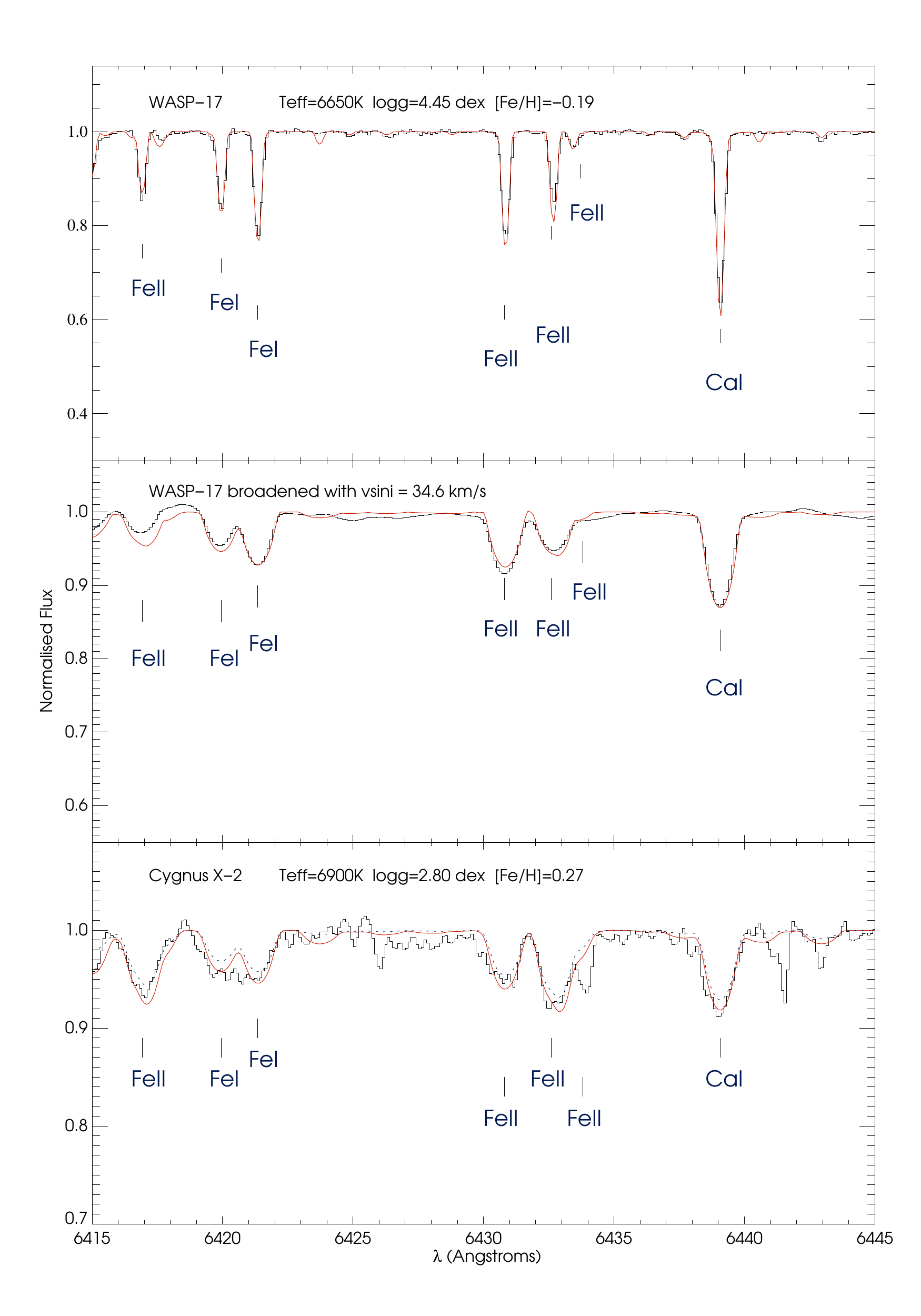}
				\includegraphics[angle=0,width=5.75cm]{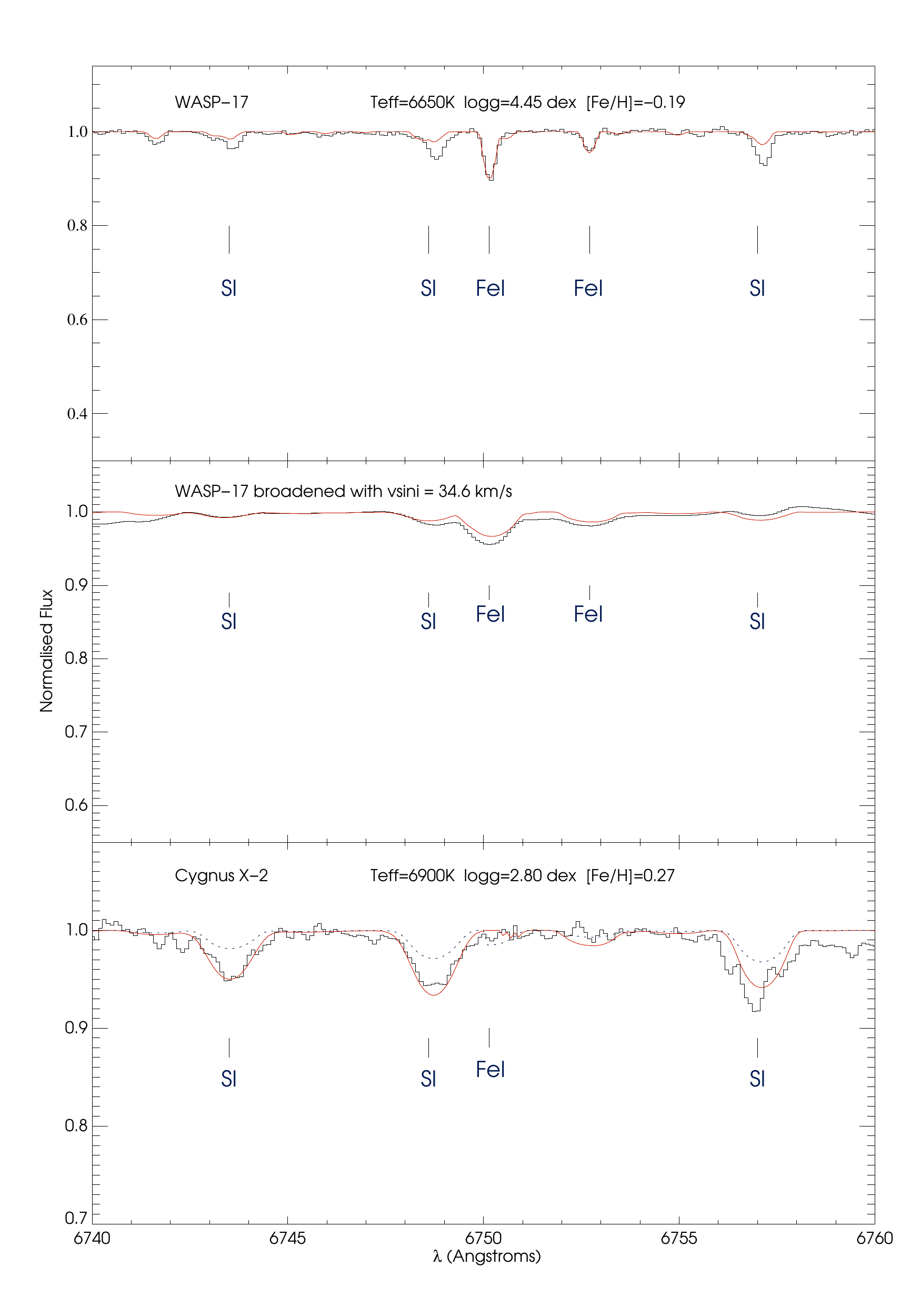}
		\hspace{1.0cm}} \caption{
		Best synthetic fit to the UES spectrum of the secondary star in the neutron star X-ray binary Cygnus X-2 
		(bottom panel) and best fit 
		to our template, with and without rotational broadening (middle and top panels). Synthetic spectra are computed for best-fit
		abundances (solid line) and for solar abundances (dashed line). Figures taken from \cite{suarez15}
	}
	\label{cygnus}       
\end{figure}

Several spectral regions containing the lines of Si, Ca, Al, Ti, Ni
have been analyzed. Although
the lines of these elements were usually the main contributor to the
features, in some cases, they were blended with Fe. The inaccuracy in
the location of the continuum caused by the blends of many weak
rotationally broadened stellar lines was one of the main sources of
error in the abundance determinations. 
Several examples of the fits to specific absorption lines are shown in Figure~\ref{xte} and 
Figure~\ref{cygnus}.
Most of the results for the LMXB systems studied to date are compiled
in Table \ref{all_data} and reviewed in the next section.

\begin{table}
\caption{Masses, velocity, stellar, veiling parameters and chemical abundances$^{\dagger}$ in LMXBs. See \cite{gonzalez11} for a definition of the parameters listed in column 1.}
\resizebox{\textwidth}{!}{
\begin{tabular}{lcccccc}
\hline\noalign{\smallskip}
	{Star} & A0620--00 & Centaurus X-4 & XTE J1118+480 & Nova Sco 94 & 
		V404 Cygni & Cygnus X-2 \\ 
		\hline\noalign{\smallskip}
	Alternative name        & V616 Mon      &  V822 Cen     &  KV UMa     & GRO J1655--40  & GS 2023+338   & V1341 Cyg\\ 
	$M_{CO, f} $ ($M_{\odot}$) & $6.61\pm0.25$ & $1.50\pm0.40$ & $8.30\pm0.28$ & $6.59\pm0.45$ & $9.00\pm0.60$ & $1.71\pm0.21$ \\
	$M_{2, f}$ ($M_{\odot}$) & $0.40\pm0.05$ & $0.30\pm0.09$ & $0.22\pm0.07$ & $2.76\pm0.33$ & $0.54\pm0.08$ & $0.58\pm0.23$\\
	$v \rm sin i$ (km $s^{-1}$) & $82\pm2$ & $44\pm3$ & $100^{+3}_{-11}$ & $86\pm4$ & $40.8\pm0.9$ & $34.6\pm0.1 $ \\
	$T_{\mathrm{eff}}$ (K)  & $4900\pm100$  & $4500\pm100$  & $4700\pm100$  & $6100\pm200$   & $4800\pm100$  & $6900\pm200$\\ 
	$\log (g/{\rm cm~s}^2)$ & $4.2\pm0.3$   & $3.9\pm0.3$   & $4.6\pm0.3$   & $3.7\pm0.2$    & $3.50\pm0.15$ & $2.80\pm0.20$\\
	$f_{4500}$              & $0.25\pm0.05$ & $1.85\pm0.10$ & $0.85\pm0.20$ & $0.15\pm0.05$  & $0.15\pm0.05$ & $1.55\pm0.15$\\
	$m_0/10^{-4}$           & $-1.4\pm0.2$  & $-7.1\pm0.3$  & $-2\pm1$      & $-1.2\pm0.3$   & $-1.3\pm0.2$  & $-2.7\pm0.4$\\
	{[O/H]}$^{\ddagger}$ & --   & --		& --		& $0.91\pm0.09$  & $0.60\pm0.19$ & $0.07\pm0.35$\\
	{[Na/H]}	        & --            & --		& --		& $0.31\pm0.26$  & $0.30\pm0.19$ & --\\
	{[Mg/H]}	        & $0.40\pm0.16$ & $0.35\pm0.17$ & $0.35\pm0.25$ & $0.48\pm0.15$  & $0.00\pm0.11$ & $0.87\pm0.24$\\
	{[Al/H]}	        & $0.40\pm0.12$ & $0.30\pm0.17$ & $0.60\pm0.20$ & $0.05\pm0.18$  & $0.38\pm0.09$ & --\\
	{[Si/H]}	        & --            & --		& $0.37\pm0.21$ & $0.58\pm0.08$  & $0.36\pm0.11$ & $0.52\pm0.22$\\
	{[S/H]}                 & --            & --		& --            & $0.66\pm0.12$  &  --    &  $0.52\pm0.24$ \\
	{[Ca/H]}	        & $0.10\pm0.20$ & $0.21\pm0.17$ & $0.15\pm0.23$ & $-0.02\pm0.14$ & $0.20\pm0.16$ & $0.27\pm0.33$\\
	{[Ti/H]}	        & $0.37\pm0.23$ & $0.40\pm0.17$ & $0.32\pm0.26$ & $0.27\pm0.22$  & $0.42\pm0.20$ & $0.59\pm0.31$\\
	{[Cr/H]}	        & --            & --            & --            & --             & $0.31\pm0.19$ & --\\
	{[Fe/H]}                & $0.14\pm0.20$ & $0.23\pm0.10$ & $0.18\pm0.17$ & $-0.11\pm0.10$ & $0.23\pm0.09$ & $0.27\pm0.19$\\
	{[Ni/H]}	        & $0.27\pm0.10$ & $0.35\pm0.10$ & $0.30\pm0.21$ & $0.00\pm0.21$	 & $0.21\pm0.19$	 & $0.52\pm0.27$ \\
	\hline\noalign{\smallskip}
	
	\label{all_data}
\end{tabular}}
$^{\dagger}$ The uncertainties on the stellar abundances 
		given in this table have been derived without taking into account the error on 
		the microturbulence. 
		
$^{\ddagger}$
		Oxygen abundances are given in NLTE. \\
	
		{References: V404 Cygni: \cite{gonzalez11}; 
				Centaurus X-4: \cite{gonzalez05}
				A0620-00: \cite{gonzalez04} 
				Nova Scorpii 1994: \cite{gonzalez08a}; 
				XTE J1118+480: \cite{gonzalez08b};
				Cyg X-2: \cite{casares10, suarez15}}
\end{table}

\subsection{Individual systems}

\subsubsection{Nova Scorpii 1994 (GRO J1655-40)}
Keck/HIRES spectrum of this system has been first studied by Israelian
et al. (2009) \cite{israelian99} who discovered that Oxygen, Sulphur
and Silicon are overabundant from 8 to 10 times compared to the Sun.
The analysis of Gonz\'alez Hern\'andez et
al. (2008a) \cite{gonzalez08a} based on the VLT/UVES high-resolution
spectra confirmed that the abundances of Al, Ca, Ti, Fe and Ni are
consistent with solar values, whereas Na, and especially O, Mg, Si and
S are significantly enhanced in comparison with the Sun and Galactic
trends of these elements. A comparison with spherically and
non-spherically symmetric supernova explosion models may provide
stringent constraints to the model parameters as mass-cut and the
explosion energy, in particular from the relative abundances of Si, S,
Ca, Ti, Fe and Ni. 

\subsubsection{A 0620-00}
It has been shown \cite{gonzalez04} that the secondary star in this system is metal-rich with [Fe/H] = 0.14 $\pm$ 0.20. Nevertheless,  the abundances of Fe, Ca, Ti, Al, and Ni are slightly higher than solar. The abundance ratios of each element with respect to Fe were compared with these ratios in late-type main
sequence metal-rich stars. Moderate anomalies for Ti, Ni, and especially Al have been found. A comparison with element yields from spherically symmetric supernova explosion models suggests that the secondary star captured part of the ejecta from a supernova that also originated the compact object in A0620-00. The observed abundances can be explained if a progenitor with a  14 $M_{\odot}$ He core exploded with a mass cut in the range 11-12.5 $M_{\odot}$, such that no significant amount of iron could escape from the collapse of the inner layers. It is very important to study abundances of  O, Si, Mg, S and C to confirm this scenario.

\subsubsection{Cen X-4}
 Abundances of Fe, Ca, Ti, Ni and Al have been obtained using VLT/UVES spectra \cite{gonzalez05}
These elements are found to have super solar abundances. Iron is enhanced too with [Fe/ H] = 0. 23 $\pm$  0. 10). Interestingly, Ti and Ni are moderately enhanced as compared with the average values of stars of similar iron content. These element abundances can be explained if the secondary star
captured a significant amount of matter ejected from a spherically symmetric supernova explosion of a 4 $M_{\odot}$  He  core progenitor and assuming solar abundances as primordial abundances in the secondary star. The kinematic properties of the system indicate that the neutron star received a natal kick velocity through an aspherical supernova and/or an asymmetric neutrino emission. The former scenario might be ruled out since the model computations cannot produce acceptable fits to the observed abundances. 

\subsubsection{V4641 Sgr}

Spectroscopic analysis of this system has been carried out by Orosz et al (2001) \cite{orosz2001}. Peculiar abundance patterns have been claimed from the analysis of
low  and high resolution spectra obtained with different instruments. These authors found that N and Ti are enhanced about 10 times compared to the Sun, 
Mg over abundance is about 5 to 7 times,  O is 3 times the solar value 
while Si is about solar. Given the physical characteristics of the companion star (mass, rotation, 
evolutionary stage)  it is impossible to understand how and why Ti is extremely  enhanced while Si is solar.  More spectral lines, better analysis and better quality spectra are needed to confirm and expand this study. Given the effective temperature of the star, NLTE studies are mandatory.

\subsubsection{XTE J1118+480}
Abundance of Mg, Al, Ca, Fe, Ni, Si and Ti have been derived using medium-resolution optical spectra of the secondary star in the high Galactic latitude black hole X-ray binary XTE J1118+480 \cite{gonzalez06}. The super solar abundances indicate that the black hole in this system formed in a supernova event, whose nucleosynthetic products could pollute the atmosphere of the secondary star, providing clues on the possible formation region of the system, either Galactic halo, thick disk, or thin disk. A grid of explosion models with different He core masses, metallicities, and geometries have been explored. Metal-poor models associated with a formation scenario in the Galactic halo provide unacceptable fits to the observed abundances, therefore rejecting a halo origin for this X-ray binary. The thick-disk scenario produces better fits, although they require substantial fallback and very efficient mixing processes between the inner layers of the explosion and the ejecta. This makes very unlikely that the system was born in the thick disk . The best agreement between the model predictions and the observed abundances is obtained for metal-rich progenitor models. In particular, non-spherically symmetric models are able to explain, without strong assumptions of extensive fallback and mixing, the observed abundances. Moreover, asymmetric mass ejection in a supernova
explosion could account for the required impulse necessary to eject the system from its formation region in the Galactic thin disk to its current halo orbit.

\subsubsection{V404 Cyg}
The atmospheric abundances of O, Na, Si, Ti, Mg, Al, Ca, Fe, and Ni have been derived using KeckI/HIRES spectra \cite{gonzalez11}.  
The abundances of  Si, Al, and Ti are slightly enhanced when comparing
with average values in thin-disk solar-type stars. The O abundance, derived from optical lines, is clearly 
enhanced in the atmosphere of the secondary star in V404 Cygni. This, together with
the peculiar velocity of this system as compared with the Galactic velocity dispersion of thin-disk
stars, suggests that the black hole formed in a supernova or hypernova explosion. 
Different supernova/hypernova models having various geometries to study possible contamination
of nucleosynthetic products in the chemical abundance pattern of the secondary star have been explored 
(e.g. \cite{nak01}). A reasonable agreement between the observed abundances and the model predictions 
has been found \cite{gonzalez11}. However,
the O abundance seems to be too high regardless of the choice of explosion energy or mass cut,
when trying to fit other element abundances. Moreover, Mg appears to be underabundant for
all explosion models, which produces Mg abundances roughly 2 times higher than the observed
value. The case of V404 Cyg is very peculiar and more studies are required to understand these observations.

\subsubsection{Cyg X-2}
\begin{figure}[h] \hbox{\hspace{-0.1cm}
		\includegraphics[angle=90,width=10cm]{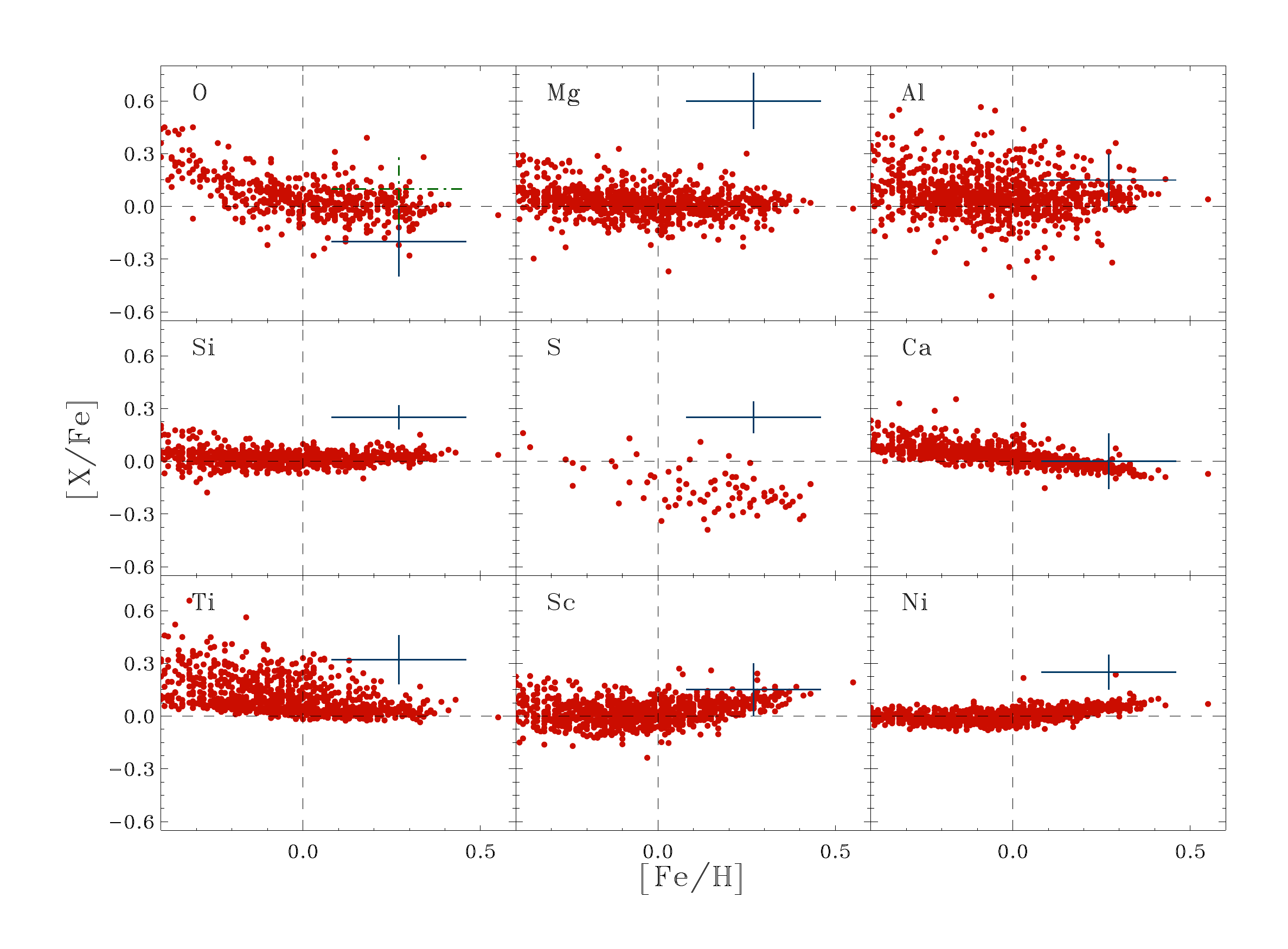}
		\hspace{1.0cm}} \caption{ 
		Abundance ratios of the secondary star in Cyg X-2 (wide cross with errors) in comparison with the abundances of solar-type metal-rich dwarf stars. Taken from 
\cite{suarez15}
		}
	\label{abun_cygnus}       
\end{figure}

Su\'arez-Andr\'es (2015) \cite{suarez15} have investigated abundances of O, Mg, Si, Ca, S,
Ti, Fe and Ni. The system is metal rich ([Fe/H]= 0, 27$\pm$0,19) and abundances of some 
some alpha-elements (Mg, Si, S, Ti) are enhanced (see Fig. \ref{cygnus}). This is consistent
with a scenario of contamination of the secondary star during the supernova event.
Strange enough,  oxygen appears to be under-abundant,  whereas 
Fe and Ni are enhanced. Assuming that these abundances come from the matter that
has been processed in the supernova and then captured by the secondary star,  
\cite{suarez15} explored 
different supernova explosion scenarios with diverse geometries. A non-spherically symmetric
supernova explosion, with a low mass cut, seems to reproduce better the observed abundance
pattern of the secondary star compared to the spherical case.
These authors have searched for anomalies in the abundance pattern of
the secondary star by comparing their results with Galactic
trends (see Fig. \ref{abun_cygnus} taken from \cite{suarez15}, and reference from that article).

As it is shown in Fig. \ref{abun_cygnus}, most of the elements in Cygnus X-2 show over-abundances when compared with Galactic trends, with the exception aluminium, calcium 
and cadmium, which are consistent with those trends.

\begin{acknowledgement}
	JC would like to thank the hospitality of the Department of Physics
	of the University of Oxford, where this work was performed during a
	sabbatical visit. 
	He also thanks Phil Charles for useful comments and discussions. 
	Finally, JC acknowledges support by DGI of the Spanish
	Ministerio de Educaci\'on, Cultura y Deporte under grants
	AYA2013-42627 and PR2015-00397 and from the Leverhulme Trust
	Visiting Professorship Grant VP2-2015-046.  
	GI thanks Luc\'\i{}a Su\'arez and Jonay Gonz\'alez Hern\'andez for useful 
	discussions. PGJ would like to thank 
	James Miller--Jones for many useful discussions and his approval to
	use data on GX~339$-$4 and Swift~J1753.0$-$0127 before their final
	publication. PGJ acknowledges funding from the European Research
	Council under ERC Consolidator Grant agreement no 647208. 

\end{acknowledgement}

\noindent
{\large{\bf Cross-Reference}}

\begin{itemize} 

\item{Chapter 7.4:  The Masses of Neutron Stars, by J. Horvath and R. Valentim}

\item{ Chapter 7.13: Supernovae and the Evolution of Close Binary Systems, by E. van den Heuvel}

\end{itemize}

%
%
%

\end{document}